\documentclass[aps,prl,twocolumn,showpacs,superscriptaddress,groupedaddress]{revtex4}  % for review and submission
\usepackage{graphicx}  % needed for figures
\usepackage{dcolumn}   % needed for some tables
\usepackage{bm}        % for math
\usepackage{amssymb}   % for math

\begin{document}

% The following information is for internal review, please remove them for submission
%\widetext
%\leftline{Made \today}
%\leftline{Primary authors: XXX}
%\leftline{To be submitted to (PRL, PRD-RC, PRD, PLB; choose one.)}
%\leftline{Comment to {\tt d0-run2eb-nnn@fnal.gov} by xxx, yyy}
%\centerline{\em D\O\ INTERNAL DOCUMENT -- NOT FOR PUBLIC DISTRIBUTION}

% the following line is for submission, including submission to the arXiv!!
%\hspace{5.2in} \mbox{Fermilab-Pub-04/xxx-E}
\author{Aurelia R. Honerkamp-Smith}
\affiliation{Department of Chemistry, University of Washington, Seattle WA 98195-1700}
\author{Benjamin B. Machta}
\affiliation{Department of Physics, Cornell University, Ithaca NY 14850}
\author{Sarah L. Keller}
\email{slkeller@chem.washington.edu}
\affiliation{Department of Chemistry, University of Washington, Seattle WA 98195-1700}

\title {Experimental observations of dynamic critical phenomena in a lipid membrane}
%\title{Dynamic critical exponent in a 2D lipid membrane with conserved order parameter}
%\input author_list.tex       % D0 authors (remove the first 3 lines
                             % of this file prior to submission, they
                             % contain a time stamp for the authorlist)
                             % (includes institutions and visitors)

\begin{abstract}
Near a critical point, the time scale of thermally-induced fluctuations diverges in a manner determined 
by the dynamic universality class. Experiments have verified predicted 3D dynamic critical exponents in
many systems, but similar experiments in 2D have been lacking for the case of conserved order parameter. Here we 
analyze time-dependent correlation functions of a quasi-2D lipid bilayer in water to show that its 
critical dynamics agree with a recently predicted universality class. In particular, the effective dynamic exponent $z_{\text{eff}}$ crosses over from $\sim\! 2$ to $\sim\! 3$ as the correlation length of fluctuations 
exceeds a hydrodynamic 
length set by the membrane and bulk viscosities.
\end{abstract}
\pacs{64.60.Ht, 68.35.Rh, 87.16.D-, 87.16.dt}

\maketitle
Lipids self-assemble in water to form sheets that are two molecules thick, within which the lipids are free to diffuse. When composed of several lipid species these two-dimensional (2D) liquid membranes can demix into coexisting liquid phases, termed $L_{\text o}$ and $L_{\text d}$, over a range of temperatures and compositions, and can exhibit critical behavior~\cite{Veatch07,Esposito07,Honerkamp08,Honerkamp09}. Among 2D critical phenomena, composition fluctuations in membranes are rather unique in that their large sizes and long decay times are accessible to optical microscopy.  For example, Fig. 1 and supplementary movies show a vesicle (a spherical membrane shell) in which correlated regions reaching $10$ $\mu$m persist for seconds~\cite{EPAPS}. Direct visualization of
these equilibrium fluctuations has recently been used to show that {\it static} critical exponents for lipid membranes are consistent with the 2D Ising universality class~\cite{Honerkamp08,Veatch08}.  Here we exploit
the ability to visualize {\it dynamics} of these fluctuations to examine for the first time the dynamic 
critical phenomena in this system.  We find that although the statics are 2D phenomena, the critical dynamics are modified by hydrodynamic coupling to the surrounding 3D fluid.
\par

Static critical exponents, which describe how observables such as correlation length 
vary as the critical point is approached, are identical for all systems in a given universality class, 
independent of their detailed microscopic physics~\cite{Goldenfeld92,Sethna06}.  For example, although membranes have a conserved order parameter and ferromagnets do not,
membranes exhibit static exponents $\nu=1.2 \pm0.2$ and $\beta=0.124 \pm 0.03$, consistent with the 
expected 2D Ising values of $\nu=1$ and $\beta= 1/8$~\cite{Honerkamp08}. Results in plasma membrane vesicles are also consistent with 2D Ising exponents $\nu= 1$ and $\gamma= 7/4$~\cite{Veatch08}.
Systems that are in the same static universality class can fall into different dynamic universality sub-classes determined by conservation laws constraining how fluctuations dissipate~\cite{Hohenberg77}.  
The critical exponent $z$ for each dynamic subclass quantitatively describes the scaling of the dynamics.  It relates how the correlation time $\tau_{s}$ diverges as temperature $T$ approaches the critical temperature $T_{c}$, such that $\tau_{s}\propto \left|(T-T_c)/T_c\right|^{-\nu z}$ where $\nu$ is the static critical exponent.  Experiments measure an effective exponent $z_{\text{eff}}$ that approaches $z$ as $T\rightarrow T_{\text c}$ and $\xi \rightarrow \infty$.
Dynamic sub-classes relevant to 2D systems with conserved order parameter are notable equally for their wealth of theoretical predictions~\cite{Hohenberg77,Haataja09,Inaura08} and for the lack of experiments that systematically test those predictions.\par

% Static properties are measurable from instantaneous configurations of an order parameter $m(\vec{r})$, whereas dynamic ones require a time series $m(\vec{r}, t)$. 

\begin{figure} [b] 
\includegraphics{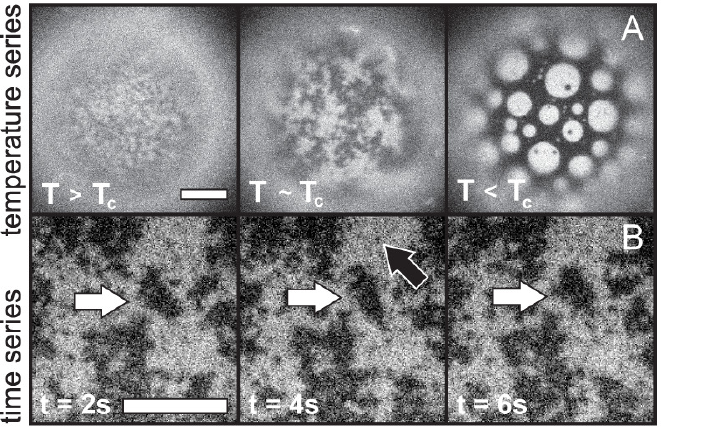}
\caption{\label{Figure 1} Fluorescence micrographs of vesicles of diameter ~200 $\mu$m. (A) As temperature changes from $T>T_{\text c}$ ($T=31.25^{\circ}\mathrm{C}$, $T_{\text c} \approx 30.9$) to $T\sim T_c$ ($T=31.0^{\circ}\mathrm{C}$) fluctuations in lipid composition grow. Below $T_c$, at $T=28^{\circ}\mathrm{C}$, domains appear. Scale bar = 10 $\mu$m. (B) A movie of composition fluctuations within a vesicle above $T_{\text c}$. Large fluctuations persist for seconds (white arrows), whereas small ones disappear by the next frame (black arrow). Scale bar = 20 $\mu$m.}
\end{figure}

Only a few previous measurements of dynamic critical exponents in 2D systems exist. Most experiments have been conducted on magnetic films. Using ferromagnetic films of $\sim$ two monolayers, Dunlavy and Venus found $\nu z = 2.09\pm0.06$, with $\nu= 1$~\cite{Dunlavy05}. Fewer experiments have been conducted on systems with conserved order parameter. Careful attempts to measure $z$ were made in thin films of lutidine and water, but were unable to reach the 2D critical regime \cite{Casalnuovo84}. In plasma membrane vesicles from living rat basophil leukemia cells, fluctuation decay times were reported to be consistent with $z \approx 2$~\cite{Veatch08}.\par

Here we obtain $z_{\text {eff}}$ as $T$ approaches $T_{\text c}$ in a lipid membrane surrounded by water and compare to theory recently developed for an analogous system: a 2D critical binary fluid embedded in a non-critical bulk fluid~\cite{Haataja09,Inaura08}. \par

This new theory incorporates three essential features of lipid bilayer dynamics: conserved order parameter, collective hydrodynamics, and hydrodynamic coupling between the bilayer and bulk~\cite{Haataja09,Inaura08}.  Inclusion of only the first feature within an Ising model yields Model B, in which composition fluctuations dissipate through diffusion of microscopic constituents~\cite{Hohenberg77}. 2D Model B predicts $z = 4-2\beta=3.75$~\cite{Hohenberg77}, and numerical schemes give $z = 3.80$ and $z = 3.95$~\cite{Yalabik82,Zheng01}. Inclusion of the first \emph{two} features, such that collective hydrodynamic motion replaces single particle diffusion as the dominant mechanism of order parameter relaxation, yields Model H.  2D Model H with coupling to only 2D momentum modes predicts $z \approx 2$~\cite{Hohenberg77}. Inclusion of all \emph{three} features yields Model HC, where HC denotes hydrodynamic coupling of the membrane to the bulk.  This new version extends Model H to account for modes in both the 2D membrane and the 3D bulk fluid, with the result that $z=3$~\cite{Haataja09,Inaura08}.\par

Intuition for the role of the coupling between the membrane and bulk within Model HC can be gleaned from an approximation for 3D Model H by Kawasaki~\cite{Kawasaki72}. Critical fluctuations are treated as spherical inclusions of diameter $\xi$ that diffuse a distance $\xi$ to equilibrate~\cite{Hohenberg77,Kawasaki72,Kadanoff68,Siggia76}. As such, correlation time varies as $\tau \sim \xi^2/D(\xi)$, where $D(\xi)$ is the inclusion's diffusion constant in a non-critical fluid. In 3D, $D(r) \sim 1/r$, where $r$ is the inclusion's radius. Using $\tau\propto\left|(T-T_c)/T_c\right|^{-\nu z}\propto \xi^{z}$ yields $z \approx 3$. A more sophisticated theoretical treatment gives $z = 3.065$~\cite{Siggia76}. Applying the same reasoning to 2D Model H, in which diffusion of inclusions has only a logarithmic dependence on $r$, yields $z \approx 2$.  Again, more sophisticated treatments produce similar values; see~\cite{EPAPS} for more detail.  This argument can be extended to predict the value $z$ should take in a 2D critical system embedded in a bulk fluid.  Classic work by Saffman and Delbr\"{u}ck examined diffusion of an inclusion in a 2D liquid of viscosity $\eta_{\text {2D}}$ immersed in a bulk fluid of 3D viscosity $\eta_{\text {3D}}$, where hydrodynamic length $L_{\text {h}}=\eta_{\text {2D}}/\eta_{\text {3D}}$ is an important parameter~\cite{Saffman75,Hughes81}. When $r \gg L_{\text {h}}$, dissipation is primarily into the bulk and $D(r)\propto1/r$ as in 3D Model H.  When $r \ll L_{\text {h}}$, dissipation is primarily into 2D hydrodynamic modes and $D(r)\propto \ln(L_{\text h}/r)$, similar to 2D Model H.  Two groups have independently noted that when $L_{\text h}$ is considered, $z_{\text {eff}}$ for a 2D critical binary fluid embedded in bulk liquid crosses over from $z_{\text {eff}} \approx 2$ when $\xi \ll L_{\text h}$ to $z_{\text {eff}} \approx 3$ when $\xi \gg L_{\text h}$~\cite{Haataja09, Inaura08}.\par

\begin{figure} [b]
\includegraphics{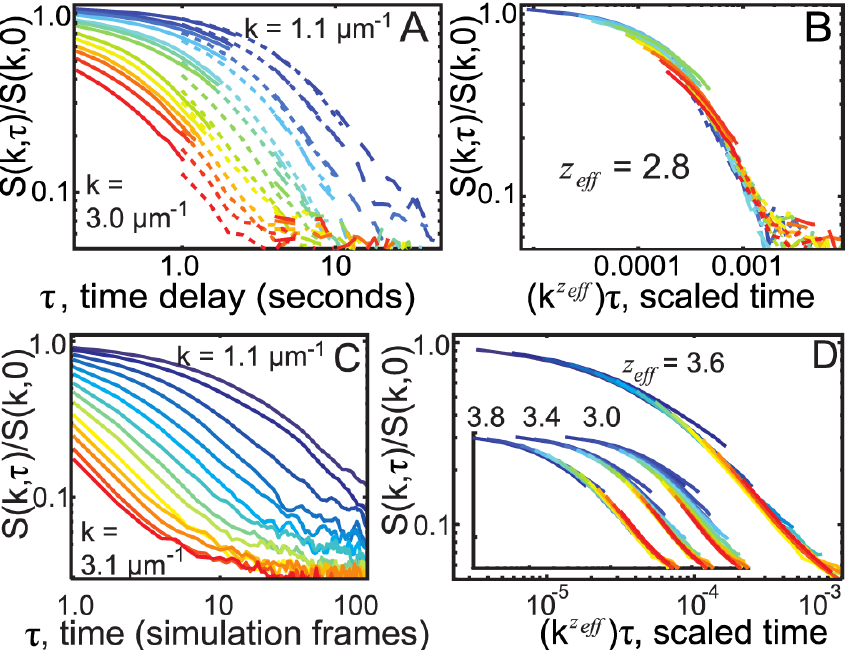}
\caption{\label{Figure2}(color online). (A and B) Rescaling experimental data closest to $T_{\text c}$ by $k^z \tau$ collapses all curves to $z_{\text {eff}}$ = 2.8, consistent with Model HC. Normalized structure factors are shown for $\xi$ = 13 $\pm$ 2.2 $\mu$m and three video rates: 10 frames per second (fps, solid lines), 2 fps (short dash), and 0.5 fps (long dash). Colors denote wavenumbers $k$ = 1.1 $\mu$m$^{-1}$ (top curve, blue) to 3.0 $\mu$m$^{-1}$ (bottom, red). (C and D) Simulations solely to verify technique. Structure factors of Kawasaki dynamics at $T = T_c$ blurred in time to mimic experimental limitations collapse at $z_{\text {eff}} = 3.6 \pm 0.2$, consistent with $z=3.75$ for 2D Model B. Colors range from $k$ = 1.1 $\mu$m$^{-1}$ to 3.1 $\mu$m$^{-1}$. Insets show collapses used to determine bounds for $z_{\text {eff}}$ and failure of collapse at $z_{\text {eff}} = 3$.}
\end{figure}

The next four paragraphs demonstrate that experimental results here are in excellent agreement with the recent predictions of Model HC, namely that $z_{eff}$ crosses over from $\sim 2$ to $\sim3$ as $T\rightarrow T_{\text c}$ and $\xi \rightarrow \infty$.  Further experimental details follow the results.

A time series of the order parameter, $m(r,t)$, was extracted from videos of vesicles collected via fluorescence microscopy.  For membranes, $m(\vec{r})$ is the deviation from average composition as reported by an image's pixel grey scales.  A time-correlation function $C(r,\tau)$, and its Fourier transform in space, the structure factor $S(k, \tau)$, were calculated for each wavenumber $k$. 

Curves of $S(k,\tau)/S(k,0)$ vs $k^{z_{\text {eff}}}\tau$ were plotted for a range of $z_{\text {eff}}$ values.  Fig. 2B illustrates how the correct $z_{\text {eff}}$ was identified: for a single value of $z_{\text {eff}}$, all experimentally-measured curves at different $k$ values (Fig. 2A) collapsed most fully onto a single curve, here at $z_{\text {eff}}=2.8 \pm 0.2$. Fig. 3A shows $z_{\text {eff}}$ values extracted in this manner from data over the entire measurable range of correlation lengths.  In Fig. 3A, $z_{\text {eff}}$ rises from from near $2$ to near $3$ as $T\rightarrow T_{\text c}$, in accord with Model HC~\cite{Haataja09, Inaura08}.\par

Fig. 2C-D validates this method by showing that standard simulations of Model B Kawaski dynamics that are blurred to mimic experimental limitations and then analyzed in the same way as the experimental data give $z = 3.6 \pm 0.2$ in agreement with the expected value of $z=3.75$ (see~\cite{EPAPS} for details). Simulations were run on a $400$x$400$ bi-periodic square lattice. Blur was achieved by averaging snapshots over 200 consecutive Monte Carlo sweeps, leaving a break of 800 sweeps without snapshots, and repeating the process, which reproduced the effects of a camera shutter opening for $100$ ms of every $500$ ms.\par

\begin{figure}[b]
\includegraphics{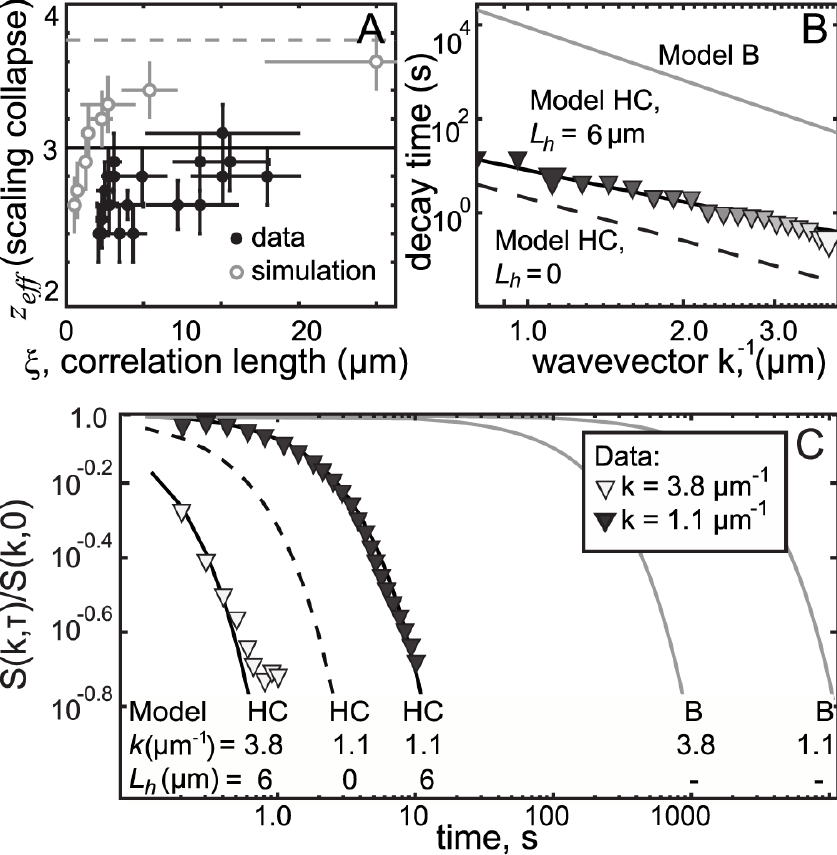}
\caption{\label{Figure3} Data is in excellent agreement with Model HC. (A) Filled symbols: Dynamic exponent $z_{\text {eff}}$ from scaling collapse of experimental data as in Fig. 2A-B.  Open symbols: Model B simulation in which $z_{\text {eff}}$ approaches $\sim 3.75$. (B) Decay time, defined as when $S(k,\tau)/S(k,0)=e^{-1}$. Large symbols indicate wavenumbers 1.1 and 3.3 $\mu$m$^{-1}$. (C) Normalized structure factors $S(k,\tau)/S(k,0)$.  In panels B and C, experimental data is denoted by symbols, 2D Model B by a grey line, Model HC (HC) with $L_{\text h}$ = 6 $\mu$m by a solid line and Model HC with $L_{\text h}$ = 0 by a dashed line.}
\end{figure}

Excellent agreement between predicted and measured structure factors provides even stronger evidence that Model HC describes critical dynamics in membranes.  Inaura and Fujitani give a prediction for the entire time-dependent structure factor $S(k,\tau)$ for Model HC, taking as input $\eta_{\text {2D}}$, $\eta_{\text {3D}}$, and a mean-field approximation for the static structure factor, $S(k,0)$~\cite{Inaura08}. The ratio $S(k,\tau)/S(k,0)$ and its decay time will be compared between theory and experiment below. A feature of $S(k,\tau)/S(k,0)$ is that it needs no correction due to the microscope's point spread function. Ratios of $S(k,\tau)/S(k,0)$ in the critical Ising model and in the mean-field approximation are similar, and do not depend strongly on correlation length, as will be shown in a future manuscript.\par

The HC model with $L_h$ = 6 $\mu$m fits the data over all experimentally accessible wavenumbers. Fig. 3C shows the ratio $S(k,\tau)/S(k,0)$ at wavenumbers 1.1 $\mu$m$^{-1}$ and 3.8 $\mu$m$^{-1}$. Fig. 3B shows decay times, defined as when $S(k,\tau)/S(k,0)=e^{-1}$. All other models are excluded.  Fig. 3A rules out 2D Model H because the measured $z_{\text {eff}}$ rises close to $3$, well above the predicted value of $2$ for 2D model H. Fig. 3B-C rules out 2D Model B because measured decay times are orders of magnitude shorter than the model predicts.\par

Developing Model HC to completely describe membranes requires determining only membrane viscosity, $\eta_{\text {2D}}$, as an input since the viscosity of water, $\eta_{\text {3D}}$, is known. Fig. 3C indicates that $\eta_{\text {2D}}$ must be nonzero. In the small $k$ limit, Inaura and Fujitani~\cite{Inaura08} predict a structure factor that depends only on $\eta_{\text {3D}}$.  This parameter-free prediction, equivalent to taking $\eta_{\text {2D}}=0$, underestimates time decays by a factor of $5-10$ (dashed curve, Fig. 3C). Setting the unknown $\eta_{\text {2D}}$ (or equivalently, $L_{\text h}$ = $\eta_{\text {2D}}/\eta_{\text {3D}}$) as a single fit parameter within Model HC over the entire measured range of $k$ yields $L_{\text h}$ = 6.0 $\pm$ 1.5 $\mu$m.  This value is within the range found by tracking diffusion of liquid domains across vesicle surfaces~\cite{Cicuta07, Petrov08} and is similar to values (2-4 $\mu$m) found by other methods, albeit for different lipid mixtures~\cite{Camley10, Dimova99}. An essentially equivalent method of finding $L_{\text h}$ is to calculate the Model HC structure factor using the formalism of Hohenberg and Halperin~\cite{Hohenberg77}, and to thereby extend Model HC to incorporate Ising rather than mean field statics.  Within experimental uncertainty, this modest change has no effect ($L_{\text h}$ = 5.5 $\pm$ 1.5 $\mu$m).  This and other extensions of Model HC, each leading to small corrections to the ratio $S(k,\tau)/S(k,0)$, will appear in a future manuscript.\par

Using lipid bilayers to measure critical exponents introduces both complexities and advantages,  which are outlined further in~\cite{EPAPS}.  
The first complexity is that the simplest bilayers that exhibit critical phenomena contain ternary lipid compositions. Strictly speaking, the ternary mixtures used here pass through isothermal critical mixing (plait) points rather than critical (col) points. A feature of 2D systems is that, unlike in 3D systems, no measurable change in critical exponents arises from the presence of a third component. Briefly, a small correction to scaling arises in systems that contain a third component at fixed composition rather than fixed chemical potential. Hence, $T_{\text c}$ changes, and many effective critical exponents are renormalized by a factor of $1/(1-\alpha)$, as discovered by Widom~\cite{Widom67} and generalized by Fisher~\cite{Fisher68}. For the 2D Ising case here, where $\alpha=0$, theory predicts only a logarithmic correction to singular behavior~\cite{Widom67}. 
The second complexity is that when $T$ is changed (as required in previous studies to find $\nu$ and $\beta$~\cite{Honerkamp08}, but not required here), a bilayer with fixed composition does not necessarily follow a path with constant $\langle m(\vec{r})\rangle$~\cite{Zollweg72}.  However, since membrane phase diagrams are relatively symmetric over the range of temperatures probed and since measured values of $\nu$ and $\beta$ were consistent with the 2D Ising model~\cite{Honerkamp08}, deviations from a path of constant $\langle m(\vec{r})\rangle$ are likely minor.  

The first advantage of using lipid bilayers is that it avoids challenges of other systems.  For example, lipid monolayers have confounding effects of dipole interactions, and the task of achieving simultaneous tunability and stability of surface pressure in a stationary monolayer is formidable. The second is that correlation lengths are large, partly because $\xi_0$ within the relation $\xi = \xi_0\left| (T-T_c)/T_c \right|^{-\nu}$ is on the order of the length of a lipid molecule rather than of an atom.  Separately, there is an advantage in using 2D (or quasi-2D) experimental systems over 3D systems. The critical region is larger in 2D liquid-liquid critical systems than in analogous 3D ones, partially due to differences between critical exponents in 2D vs 3D Ising classes ($\nu=1$ and $\beta=1/8$ in 2D vs $\nu\approx0.630$ and $\beta \approx 0.325$ in 3D~\cite{Goldenfeld92}). 
\par

Methods used to produce the results in Fig. 3 follow. To optimize movie quality, vesicles were spherical, free-floating, unilamellar, of radius $>$100 $\mu$m, and electroformed by standard methods detailed in~\cite{Honerkamp08}. Vesicles were formed from mixtures along a line of plait points centered at $30\%$ diphytanoylphosphatidylcholine (DiPhyPC), $20\%$ dipalmitoylphosphatidylcholine (DPPC) and $50\%$ cholesterol (chol), with $0.5\%$ fluorescent dye Texas red dipalmitoylphosphatidylethanolamine (TR-DPPE). Only vesicles near a plait point were analyzed, identified by micron-scale composition fluctuations visible over the largest observed range of temperatures ($>1^{\circ}\mathrm{C}$) and by equal areas of coexisting liquid phases below $T_{\text c}$. Each vesicle analyzed fell on a slightly different plait point, so each had a slightly different $T_{\text c}$~\cite{Veatch05}.\par

Images of membranes were captured via an epifluorescence microscope with a temperature-controlled stage and a mercury lamp source. Light exposure was minimized by employing a SmartShutter (Sutter Instrument, Novato CA) controlled through NIS-Elements (Nikon, Melville NY) and by recording movies for at least two different frame rates at each temperature. Each frame was exposed 100-150 ms, with the shutter open 10 ms before and after exposures. Movies were collected from high to low temperature in steps of $\sim0.2^{\circ}\mathrm{C}$, equilibrated for at least 2 min.  No consistent trend in intensity was observed throughout each movie, implying that the low light procedures used here eliminated significant photobleaching. To correct for lamp flickering, mean brightness was subtracted from each frame. Spatial intensity gradients due to other vesicles outside the focal plane were removed by a long wavelength filter of 100 pixels.\par 

Images were analyzed via custom MATLAB code (The Mathworks, Natick, MA). Vesicles were tracked and centered to remove drift (typically $<25\mu m$/min). By eye, features exhibit no net translation, which implies no significant vesicle rolling. No difference in mean intensity or noise between pixels at edges vs. centers of cropped images was observed, implying that vesicles are so large that membrane curvature over images can be neglected~\cite{Honerkamp08}. Curvature corrections in smaller vesicles were minor~\cite{Veatch08}.\par

The structure factor $S(k, \tau)$, the Fourier transform in space of the time-dependent correlation function, was found as previously described~\cite{Honerkamp08,Takacs08}. Briefly, a discrete transform was performed for each movie image, with a buffer of zero values to correct for image non-periodicity. Transformed images were divided by the microscope's finite point spread function to yield $m(\vec{k},t)$. The dynamic structure factor was generated at each $\tau$ by $S(\vec{k},\tau)=1/2\left\langle m(\vec{k},t)\overline{m(\vec{k},t \pm \tau)}\right\rangle$, where $\overline{m(\vec{k},t)}$ is the complex conjugate of $m(\vec{k},t)$~\cite{Kolin06}.  $S(\vec{k},\tau)$ was then radially averaged to yield $S(k,\tau)$.\par

Structure factors were employed in two ways. First, correlation lengths, $\xi$, were found by analyzing structure factors at $\tau=0$. Specifically, a one-parameter fit for $\xi$ was made until all data for $k^{(7/4)}S(k)$ vs. $k\xi$ collapsed onto the single curve for the exact numerical solution of the 2D Ising model~\cite{Wu76, Honerkamp08}. Second, effective dynamic scaling exponents, $z_{\text {eff}}$, were found by collapsing curves of $S(k,\tau)$ (see results above and~\cite{EPAPS} for details). Collapse works because, according to the dynamic scaling hypothesis, structure factors within the scaling regime can be written in the form $S(k,\tau,\xi) = k^{-2+\eta}\Omega((k\xi)^{-1},k^z\tau)$ where $\Omega$ is a universal function of $(k\xi)^{-1}$ and $k^z\tau$~\cite{Hohenberg77}.  Near $T_c$, where $(k\xi)^{-1}$ is near $0$, curves of $S(k,\tau)/S(k,0)$ vs $k^z\tau$ collected over many wavenumbers $k$ should collapse via a one-parameter fit to produce the correct value of $z$.  Here, $\Omega$ can also depend on $kL_{\text h}$, so that $S(k,\tau, \xi)=k^{-2+\eta}\Omega((k\xi)^{-1},k^z\tau, kL_{\text h})$.  For collapses in Fig. 2A-B, $z_{\text {eff}}$ refers to an effective $z$ value which varies as $\xi/L_{\text h}$ is changed. In Fig. 3B-C, comparing the entire form of the structure factor to theoretical predictions directly verifies the value of $z$ as well as the dependence of the universal function on $kL_{\text h}$ and $k^z\tau$.

\textit{Summary:} Directly imaging composition fluctuations enables measurement of effective dynamic critical exponents of a lipid membrane embedded in bulk water.  Experimental structure factors are in excellent agreement with an emerging theoretical prediction in which 3D hydrodynamics affects critical slowing down in a 2D membrane.  The theory invokes hydrodynamic coupling between the membrane and bulk fluid such that Ising degrees of freedom are coupled to momentum modes~\cite{Haataja09,Inaura08}.  As predicted, a shift in $z_{\text {eff}}$ from $\sim 2$ to $\sim3$ as $T\rightarrow T_{\text c}$ and $\xi \rightarrow \infty$ is observed. \par

This work was supported by the NSF (MCB-0744852), a Molecular Biophysics Training Award (NIH 5 T32 GM08268-20), a UW Center for Nanotechno-logy IGERT (DGE-0504573), and NIH k99GM-087810. P. Cicuta kindly provided the domain-tracking Matlab routine~\cite{Honerkamp08} customized here. J.R. Ashcraft, M.E. Cates, R.E. Goldstein, M. Haataja, T. Lubensky, D.R. Nelson, M. den Nijs, P.D. Olmsted, G. Garb$\grave{e}$s Putzel, M. Schick, J.V. Sengers, J.P. Sethna, S.L. Veatch, B. Widom and A. Yethiraj are thanked for insightful conversations.

%\bibliographystyle{prsty2}
%\bibliography{PRLbiblioMarch12AHS}

\newpage

%\title{Critical Hydrodynamics of Lipid Membranes}
%XXXproposed title change?, uncomment previous line to switch..
%\input author_list.tex       % D0 authors (remove the first 3 lines
                             % of this file prior to submission, they
                             % contain a time stamp for the authorlist)
                             % (includes institutions and visitors)
%\pacs{64.60.Ht, 68.35.Rh, 87.16.D-, 87.16.dt}

\section{Supplementary Information}

\subsection{Introduction}
This supplemental section contains details regarding analysis, calculation, and theory not found in the main text.   Section~\ref{sec:Exp} contains additional information on analysis procedures.  Section~\ref{sec:alpha} explains why Fisher renormalization does not affect the observed static critical exponents in 2D Ising systems, and why the critical region is larger in 2D liquid-liquid membrane systems than it would be in analogous 3D ones. Section~\ref{sec:sim} details the Ising model simulations that we used to verify our analysis. Section~\ref{sec:2DH} reviews previous theoretical results for purely 2D critical hydrodynamics.  

\subsection{Details of Analysis\label{sec:Exp}}
In the definition of $S(\vec{k},\tau)=\left\langle m(\vec{k},t)\overline{m(\vec{k},t + \tau)}\right\rangle$, the value of $S(k,\tau=0)$ is guaranteed to be real as an expectation value and also from run to run since each term appears with its complex conjugate.  For $\tau \neq 0$, the expectation value of the imaginary part of $S$ is guaranteed to be real by time-reversal invariance expected of systems in equilibrium.  However, each term contributes an imaginary component.  Adding the complex conjugate and dividing by two leads to an effective measurement of $1/2(S(k,\tau)+S(k,-\tau))$, which for an equilibrium system is equal to $S(k,\tau)$.   The inclusion of zeros required to pad the raw data in $m(k,t)$ introduces a small error in the calculation of $S(k,\tau)$, which can be corrected for by dividing $S(k,\tau)$ by the correlation function of pure ones and zeros in the real space correlation function.  For calculations in $k$-space, there is no simple correction (the real space correction has an ill-posed Fourier transform and so introduces unacceptable noise in $k$-space).  Nevertheless, any correction is expected to be small (values of $S(k,\tau)$ were similar when calculated with vs without zero padding).  More importantly, any correction would cancel out of the main results presented here, where $S(k,\tau)$ is divided by $S(k, \tau=0)$. For movies at the slowest frame rate, 0.5 fps, noise in $S(k,0)$ caused an offset from the rest of the structure factor.  For calculations made with those data sets, the measured value of $S(k,0)$ was replaced by the value extrapolated from an exponential fit to the 2nd through 5th points in $S(k,\tau)$. \par

\subsection{Experimental advantages of lipid bilayers~\label{sec:alpha}}  
This section explains why Fisher renormalization does not affect the observed static critical exponents in 2D Ising systems, and why the critical region is larger in 2D liquid-liquid membrane systems than it would be in analogous 3D ones.

\textit{Widom-Fisher rescaling leads to only an immeasurably small correction to 2D critical exponents}:
Concentration fluctuations consistent with static 2D Ising critical exponents were previously observed in membranes over our entire range of $k\xi$~\cite{Honerkamp08bib}.  As we show below, this observation is not at odds with the fact that our ternary system is subject to rescalings first discovered by Widom~\cite{Widom67} and generalized by Fisher~\cite{Fisher68}.  Rescaling corrections apply to any system with a quantity whose chemical potential smoothly affects the critical temperature.  When a component is instead held at fixed composition (as our three components are) then the observed critical behavior receives non-analytic corrections, essentially because the chemical potential of the third component has singular behavior near the critical point when held at fixed composition.  As a result, the singular form of the coexistence curve near the critical point is changed from its usual exponent $\beta$ to $\beta^\prime=\beta/(1-\alpha)$.  Here $\alpha$ is the static critical exponent for specific heat.  Other critical exponents that relate singular behavior of a quantity to the distance in temperature from the fixed point (for example $\alpha$, $\beta$, $\nu$ and $\gamma$) receive similar corrections.  For example, the specific heat exponent itself becomes $\alpha^\prime=\alpha/(1-\alpha)$\cite{Fisher68}.  The rescaling correction is not confined to ternary systems:  binary systems held at fixed density, rather than fixed pressure undergo similar rescalings, as does any system in which a density variable is held fixed in a phase diagram rather than it's conjugate field.\par

Fortunately for the current study, Widom-Fisher rescaling in the 2D Ising model leads to an immeasurably small change in the singular behavior and no change in the critical exponents themselves. As noted above, $\alpha$ is the static critical exponent for specific heat, $C$, which diverges as $C \sim ((T-T_c)/T_c)^{-\alpha}$ .  In the 2D Ising model, specific heat diverges as $C \sim \log((T-T_c)/T_c)$, which is slower than any power law divergence, so that $\alpha$ is said to be zero.  As explicitly discussed in both~\cite{Widom67} and~\cite{Fisher68}, there is potentially a logarithmic correction to the singular behavior of quantities whose critical exponents are usually multiplied by $(1-\alpha)^{-1}$.  For example, the correlation length $\xi$, which is usually written as $\xi \propto (\frac{T-T_{\text c}}{T_{\text c}})^{-\nu}$, becomes $\xi \propto (\frac{-\log{(T-T_{\text c})}(T-T_{\text c})}{T_{\text c}})^{-\nu}$after rescaling.  In this case, $\nu^\prime=\nu$, so rescaling does not change the critical exponent.  For a system described by the 3D Ising model (rather than the 2D Ising model as in the current study), $\alpha \approx 0.11$~\cite{Fisher68, Chaikin95}, such that the effect of rescaling is small but observable in the critical exponents. \par 

Although the effect of rescaling on the dynamic exponent $z$ is not discussed explicitly in the literature, we expect that $z$ would not be affected by rescaling even for systems in which $\alpha \neq 0$.  The dynamic exponent $z$ describes scaling of the time scale as the length scale is changed, with $\tau_{s} \propto \xi^{z}$.  The product $\nu z$ describes critical slowing with respect to temperature, where $\tau_{s} \propto  (\frac{T-T_{\text c}}{T_{\text c}})^{\nu z}$.  As such, $\nu z$ does describe the singular behavior of a quantity (here the time scale) as temperature is changed and it receives a correction of $\frac{1}{1-\alpha}$ through the parameter $\nu$. \par

\textit{The critical region is larger in 2D liquid-liquid membrane systems than in analogous 3D ones.}
The point here goes beyond considerations that as dimension decreases, the critical region becomes larger due to changes in the Ginzburg temperature, which is the temperature at which a system crosses over from mean-field to critical behavior~\cite{Chaikin95}. Correlation lengths of $\xi \sim 10\mu m$ are regularly observed in vesicle membranes.  These correlation lengths are larger than those typically observed in 3D binary mixtures, even though control over lipid composition in membranes is coarser.  This is because:
(1) lipids are molecules with length scales of $\xi_0 \sim 1nm$, whereas the atoms employed in many studies of 3D critical phenomena are an order of magnitude smaller, with $\xi_0 \sim 0.1nm$, and
(2) differences between critical exponents in the Ising classes are favorable to 2D systems.  In 2D, $\nu=1$ and $\beta=1/8$, whereas in 3D, $\nu\approx0.630$ and $\beta \approx 0.325$~\cite{Goldenfeld92}).
The scaling form for correlation length can be written as $\xi=\xi_0t^{-\nu}\mathcal{U}\left(\left(\left|\phi\right|/\phi_0\right)^{1/\beta}t^{-1}\right)$, where $t=\left(T-T_{\text c}\right)/T_{\text c}$ is reduced temperature, $\xi_0$ is a  molecular scale, and `magnetization' $\phi$ is a function of
composition normalized by $\phi_0$, which is roughly the difference in composition between the two low temperature phases far from the critical point.  $\mathcal{U}(x)$ is a universal function of its dimensionless argument $x$. $\mathcal{U}(x)$ has a maximal value at $x=0$ and decreases to 0 as
its argument increases.   Tuning a system's correlation length to $\xi$, requires tuning temperature to within approximately $\Delta T \sim T_{\text c} (\xi/\xi_0)^{-1/\nu}$.  Consider a membrane with values observed here: $\xi \sim$ 10 $\mu$m, $\xi_0 \sim$ 1 nm, and  $T_{\text c} \sim$ 300K. In
2D, $\Delta T  \sim .03$ K, which is experimentally achievable.  For 3D systems in which $\xi_0$ is typically 1 $\AA$, a $10\mu$m correlation length would require tuning temperature to a much higher precision of $\Delta T  \sim$ 10$^{-4}$K.  Similarly, extreme accuracy in composition is not required in 2D to observe fluctuations of $\xi \sim$ 10 $\mu$m. For a membrane to lie in the critical region, composition must be tuned such that the argument of $\mathcal{U}$ is of order $1$.  Given the difference between critical exponents in 2D vs 3D, far less precise experimental control of composition can be tolerated in 2D than in 3D.

\subsection{Simulation details \label{sec:sim}}
Simulation procedures were standard~\cite{Newman99,Sethna06} and briefly explained here. The standard Ising Hamiltonian given by $H=-\sum_{\left\{i,j\right\}}s_i s_j$ was used, with spin variables $s_i=\pm1$ and summation over the four nearest neighbors ($j$) of every state ($i$).  Temperatures were in terms of the exact critical temperature given by the Onsager solution~\cite{Plischke06}, $T_{\text c}=2/\log(1+\sqrt{2})$ so that a reduced temperature $t=(T-T_{\text c})/T_{\text c}$ corresponds to a simulation temperature of $T_{sim}=2.269(1+t)$.  In this section, $T$ and $\Delta H$ correspond to temperature and to the change in energy between initial and final states, respectively.  Both are in dimensionless units. In a Monte-Carlo 'sweep', $160,000$ ($400^2$) pairs of spins were proposed to be swapped, such that each spin was proposed twice.  Metropolis spin exchanges were used; each pair was exchanged or not to satisfy detailed balance~\cite{Newman99,Sethna06}.  If the resulting configuration was lower in energy, the exchange was accepted.  If energy increased, the exchange was accepted stochastically with probability $\exp(-\Delta H/T)$.\par

Note that any dynamics that satisfy detailed balance will lead to the same equilibrium ensemble of configurations~\cite{Sethna06}.  To rapidly equilibrate the system, 'nonlocal' moves were employed in which each of a pair of spins were chosen from all sites on the lattice.  Equilibration is very rapid using these nonlocal dynamics since they approximate "Model A" for large systems where $z$ is near $2$~\cite{Hohenberg77}. The system was equilibrated for $100,000$ sweeps using nonlocal moves starting from a distribution that contained the desired fraction of up spins but was otherwise random.  $100,000$ sweeps is much longer than the decay time of the slowest decaying system used here.  The decay time is approximately $1000$ sweeps at 1.05$T_{\text c}$, which can be seen qualitatively by inspecting successive snapshots or quantitatively by inspecting the decay of time dependent correlation functions. Once the system was equilibrated, dynamics relevant for the locally conserved order parameter (Kawasaki Dynamics) were employed.  In this case, a single spin and one of its four nearest neighbors were chosen to form a pair proposed to be swapped.\par

\subsection{Predictions for binary liquids in 2D~\label{sec:2DH}}  Model H for binary fluids in 2D predicts $z \approx 2$ using $z=4-\eta-x_\lambda$, where $\eta=2\beta$ is a static critical exponent and $x_\lambda$ must be calculated from an epsilon expansion (where $\epsilon=4-D$, and $D$ is the number of dimensions). This yields $x_\lambda=18/19(1-\text{(constant)}\epsilon +\mathcal{O}\epsilon^2)$ where the constant is either $0.033$~\cite{Siggia76} or $0.039$~\cite{Hao05}. Since the constant is small, it is plausible that the expansion applies even when $\epsilon= 2$, yielding $z = 2.00$ (which also arises from a much simpler mean field argument) or $1.98$.  Simulations in 2D binary liquids are reportedly challenging and we know of none that either verify or contradict the prediction that $z = 2$. Measurements in bulk 3D liquids far from $T_c$ find that $z_{\text {eff}} = 2$ (e.g.~\cite{Lastovka66,Swinney73,Burstyn82}).
 \par

%\bibliographystyle{prsty2}
%\bibliography{PRLbiblioMarch12AHS}

\end{document}